\documentclass[conference,letterpaper]{IEEEtran}

\usepackage[utf8]{inputenc} 
\usepackage[T1]{fontenc}
\usepackage{url}
\usepackage{ifthen}
\usepackage{cite}
\usepackage[cmex10]{amsmath} 
\usepackage{url}
\usepackage{graphicx}        
\usepackage{wrapfig}
\usepackage{multicol}        
\usepackage[bottom]{footmisc}
\usepackage{subfigure}
\usepackage{amsfonts}
\usepackage{amssymb}
\usepackage{epstopdf}
\usepackage{braket}
\usepackage{comment}

\DeclareMathOperator*{\argmin}{arg\,min}

\DeclareMathOperator*{\KL}{KL_\mathcal{B}}

\begin{document}
\title{Minimizing costs of communication with random constant weight codes} 


\author{%
  \IEEEauthorblockN{ Pau Vilimelis Aceituno}
  \IEEEauthorblockA{
                    Institute of Neuroinformatics\\
                    ETH Z\"urich\\ 
                    5083, Z\"urich\\
                    Email:  pau@ini.uzh.ch}
}

\maketitle

\begin{abstract}
We present a framework for minimizing costs in constant weight codes while maintaining a certain amount of differentiable codewords. Our calculations are based on a combinatorial view of constant weight codes and relay on simple approximations.
\end{abstract}


\section{Introduction}
In classical information theory, the cost of sending messages is the length of the codewords used \cite{InfThClassic}. This notion of cost is very useful when information is sent sequentially and when the cost of sending a bit is independent of the bit's value. However, in some modern applications the cost of transmitting one of the symbols is larger than the other and we need to transfer this information in blocks of a given length. For example, in neuromorphic systems the activation of a unit is more costly than having that unit remain inactive, while realistic applications require immediate transmission \cite{BenchmarksNeuromorph}. Those constraints restrict codewords to a predefined number of units, hence a fixed codelength. In this paper we present a framework to optimize the parameters of such codes. 

In the first section of this paper we present the basic equations that allow us to minimize the cost of the code while maintaining a minimum amount of distinct codewords. In the second section we address the problem of having noise and show how to compute the probability of decoding the wrong codeword given the hamming distances between codewords. In the last section we derive the expected hamming distances for random fixed weight codes and obtain the condition that guaranteed error-free codeword transmission, which we can apply to our cost minimization equation.

\section{Encoding a source}
\subsection{Problem Statement}
We start with a well-known framework: a source, an encoder a channel and a decoder. The source generates symbols probabilisticaly that are later passed onto the encoder, which converts these symbols into codewords to be transmitted through the channel and received by the decoder that finds the original symbol. The key problem that we are addressing is the issue of designing an encoder under the following assumptions:
\begin{enumerate}
	\item Units have two states, 0 or inactive, and 1 or active. 
	\item The cost function is monotonic on the number of units and active units.
	\item The number of active units is fixed. 
\end{enumerate}
We will also assume that the source generates symbols with equal probability, a requirement that is easily fulfilled if we assume that the encoder can concatenate long chains of symbols, creating composite symbols whose probability distribution follows the asymptotic equipartition property.

Our first problem is how to choose the number of units and active units as to minimize a cost function $C(a,N)$ under the constraint that the number of codewords must be larger than the number of symbols to be covered $|\mathcal{W}|$. Given that the number of codewords available for an activation of $a$ out of $N$ units is given by ${N \choose a}$, this gives us the constraint,
\begin{equation}\label{eq:constraintCodewordCount}
	{N \choose a} \geq |\mathcal{W}|.
\end{equation}

Together this can be expressed as
\begin{equation}\label{eq:costMinNoiseless_0}
	\argmin_{r,N} C(rN,N) \quad s.t.\ \ln\left(|\mathcal{W}|\right) \leq N H(r),
\end{equation}

This problem and its solution can be visualized in Fig. 1.
\begin{figure*}[h!]
	\centering
	\includegraphics[width=1\linewidth, trim={0cm 10cm 0cm 0cm},clip]{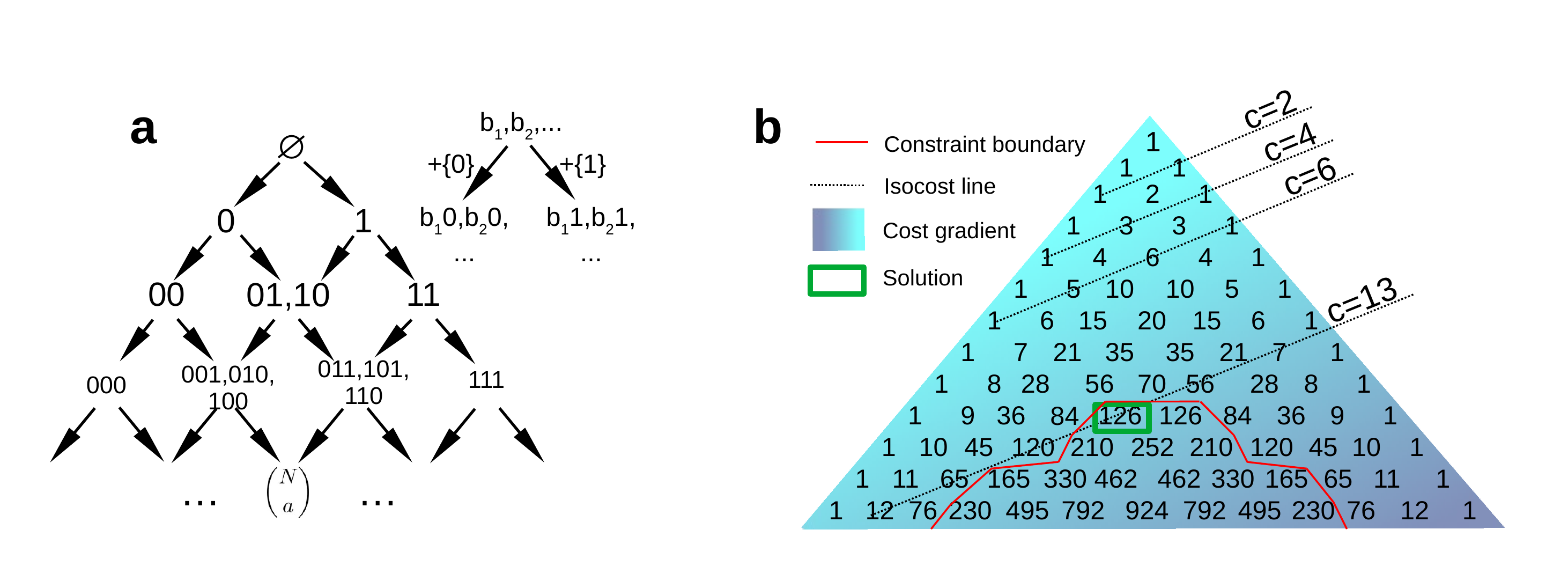}
	\caption{Graphical explanation of minimizing the cost $C(a,N) = a+N$ subject to $\mathcal{W}= 125$: The constraint in Eq.~\ref{eq:constraintCodewordCount} is presented in \textbf{a}: we start by having no units on the top level, each level down adds one unit, the arrows towards the left add a zero to the set of symbols in the original node while the arrows towards the right add a one. This tree defines Pascal's triangle, which places ${N \choose a}$ codewords on the $N$th level and the $a$th leftmost place. On this triangle we place the constraint boundary (red) as well as the gradient of the cost (blue) and the isocost lines (black dotted), with the lowest cost being at $N=9,\ a=4,\ C=13$. Note that although simple, a linear cost function on $a$ and $N$ does appear in real neuromorphic systems \cite{charlotteEnergy}.}
	\label{fig:minimalCost}   
\end{figure*}  

\subsection{Analytic Solution}
Even though the the problem is conceptually simple, the fact that the parameters are integers with combinatorial functions makes it difficult to solve it in an efficient manner. We will now present an approximate solution that can be solved efficiently, although it only gives the right order of magnitude rather than precise values of $a,N$. By using Stirling's approximation as shown in App.~\ref{App:CombToExp}.1 and defining $r = \frac{a}{N}$, 
\begin{equation}
	\begin{aligned}
		\ln{N \choose Nr} 
		&\sim N\left[r\ln\left(r\right) + (1-r)\ln\left((1-r)\right)\right] = N H(r) 
	\end{aligned}
\end{equation}
where $H(\cdot)$ is the entropy function.

The problem of choosing the right values of $a,N$ is then 
\begin{equation}\label{eq:costMinNoiseless}
	\argmin_{r,N} C(rN,N) \quad s.t.\ \ln\left(|\mathcal{W}|\right) \leq N H(r),
\end{equation}
which we can solve by noting that 
\begin{equation}
	N\geq \dfrac{\ln\left(|\mathcal{W}|\right)}{H(r)}
\end{equation}
and since we assume that the cost increases monotonically with $N$, we can solve the continuous version of our previous problem by
\begin{equation}\label{eq:solveCostMinNoiseless}
	r^* = \argmin_{r}\ C\left(r\frac{\ln\left(|\mathcal{W}|\right)}{H(r)}, \frac{\ln\left(|\mathcal{W}|\right)}{H(r)}\right), \quad N^* = \dfrac{\ln\left(|\mathcal{W}|\right)}{H(r^*)},
\end{equation}
where the first equation can be solved numerically and its solution applied to the second one.

\section{Information transmission through a noisy channel}
The subsequent problem is how to handle probabilistic errors which might corrupt a sent codeword and make the decoder confuse it with another. Our derivations are similar to the ones presented in \cite{PhyInfComp} to study Shannon Codes, which are in turn based on the notion of a distance enumerator presented in detail in \cite{InfTh}.

The first thing to do is set the basic nomenclature. An encoder sends a codeword $w$, which is transmitted and corrupted by noise. These errors will change the distance between the received codeword $w_{r}$ and the originally sent one $w_{s}$. The decoding will then be erroneous if there is another codewords that is closer to the $w_{r}$ than $w_{s}$. The original codebook $\mathcal{W}$ contains all possible codewords and we will denote the non sent ones as $w_{\times} \in \mathcal{W}-\left\{w_{s}\right\}$.

\begin{figure}[h!]
	\centering
	\includegraphics[width=1\linewidth, trim={0cm 24cm 3cm 1.5cm},clip]{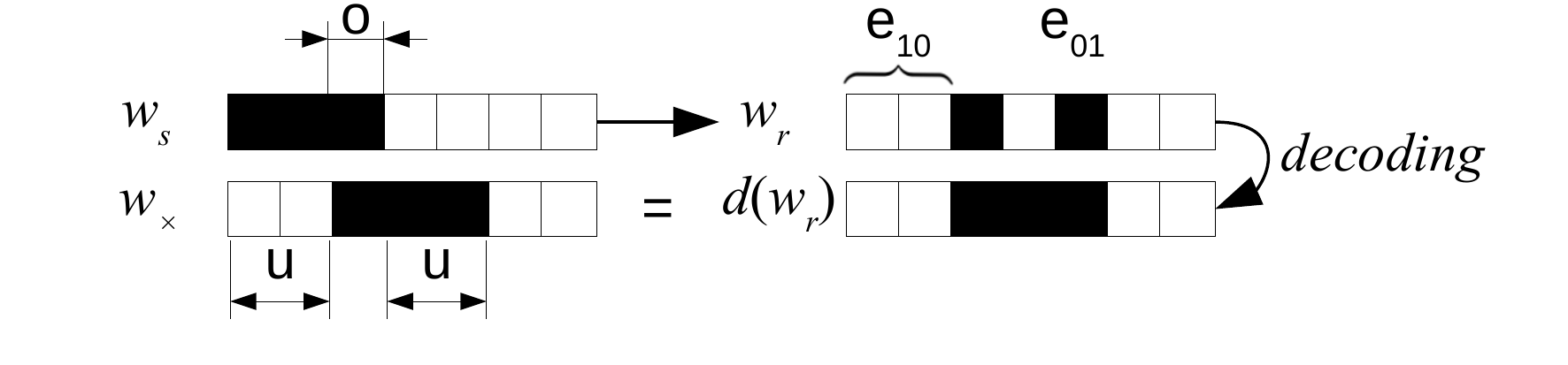}
	\caption{Illustration of the incorrect decoding process: In a dictionary with $N=7,\ a=3$ there are two codewords, $w_{s}$ and $w_{\times}$, which overlap in $o$ units and differ in $u$ units, $w_{s}$ is transmitted, noise induces $e_{10}=2,\ e_{01}=1$. Those errors affect the decoding process, because the received codeword $w_{r}$ is closer to $w_{\times}$ than to the original $w_{s}$.}
	\label{fig:overlap}   
\end{figure}

The next step is to compute the probability that a codeword corrupted by noise will be mistakenly decoded as $w_{\times}$ given an overlap $o$ between $w_{s}$ and $w_{\times}$. The decoding depends on the overlap between possible codewords, yielding an error in decoding if
\begin{equation}
	d(w_{r})\neq w_{s} \iff \exists w_{\times} \ s.t.\ \langle w_{r}, w_{s} \rangle \leq \langle w_{r}, w_{\times}\rangle = a-2u,
\end{equation} 
where $\langle\cdot, \cdot\rangle$ is the amount of active units that overlap between two codewords and $2u$ is the hamming distance between the two. 
When two words have an overlap of $o$, each one has $u = a-o$ bits that do not overlap with the other. The wrong decoding happens if enough bits of the non-overlapping sets are changed. There are two types of error to consider,
\begin{itemize}
	\item Errors $1\rightarrow0$ are denoted $e_{10}$ and appear in $u$ non-overlapping, active bits of $w_{s}$ decreases $\langle w_{r}, w_{s} \rangle$ by one.
	\item Errors $1\rightarrow0$ are denoted $e_{10}$ and appear in $u$ non-overlapping, active bits of $w_{s}$ increases $\langle w_{r}, w_{\times} \rangle$ by one.
\end{itemize}
The sum of both errors must compensate the original hamming distance between the two words. These quantities are illustrated in Fig.~\ref{fig:overlap}. Given that this distance is $2u$, a mistake requires $u$ (or more) errors of either kind. The probability of the decoding operation $d(\cdot)$ returning a random codeword that overlaps in $o$ bits is then
\begin{equation}\label{eq:DecodingErrorGivenOverlapComb}
	\begin{aligned}
		\text{Pr}&\left[d(w_{r})= w_{\times}|u\right]
		= \text{Pr}\left[\langle w_{r}, w_{s} \rangle \leq \langle w_{r}, w_{\times}\rangle| u\right]\\
		&= \text{Pr}\left[j = e_{10} \right] \text{Pr}\left[i = e_{01} \right]\Theta\left(j+i\geq u\right)\\
		&=\sum_{i=0}^{u} \sum_{j=0}^{u}{u \choose j}p_{10}^j(q_{10})^{u-j} {u \choose i}p_{01}^{i}(q_{01})^{u-i}\Theta\left(j+k\geq u\right)
	\end{aligned}
\end{equation}
where $q_{10}=1-p_{10},\ p_{01}=1-q_{01}$. Note that knowing $a$ and $o$ is equivalent to knowing $u$. This can be converted into the integral of an exponential (see App.~\ref{App:CombToExp}.2), 
\begin{equation}\label{eq:DecodingErrorGivenOverlap}
	\begin{aligned}
		\text{Pr}&\left[d(w_{r})= w_{\times}|o\right]\\
		&\sim \int_{\tiny
			\begin{array}{l}
				x,y\in \left[0,1\right]\\
				x+y \geq 1
		\end{array}}
		\exp\left\{-u\left[\KL\left(x,p_{10}\right)+\KL\left(y,p_{01}\right)\right]\right\}  dx dy
	\end{aligned}
\end{equation}
where $\KL$ is the Kullback-Leiber divergence between two Bernoulli distributions and the domain of integration is $\left\{x,y \ | 0\leq x \leq 1, 0\leq y \leq 1,  x+y \geq 1\right\}$. Notice that this integral can be approximated when $u\rightarrow\infty$ by using Laplace's method (App.~\ref{App:LaplaceExpand}). We will thus note 
\begin{equation}
	\text{Pr}\left[d(w_{r})= w_{\times}|o\right] \sim \exp\left\{-u I(p_{10}, p_{01})\right\}
\end{equation}
where $I(p_{10}, p_{01})$ is the maximum of $\KL\left(x,p_{10}\right) + \KL\left(y,p_{01}\right)$ in the domain of integration.

Given the probability that $w_{r}$ will be decoded as $w_{\times}$, we can bound the probability that there will be an error when many codewords are present. The approach is to count how many codewords have an overlap of value $o$ and then combine both values to get the expected number of wrong codewords that are closer to $w_{r}$ than $w_{s}$. This expectation is then an upper bound on the probability of making a mistake by Markov's inequality
\begin{equation}
	\begin{aligned}
		p_{\times} =&\text{Pr}\left[ d(w_{r})\neq w_s \right] = \text{Pr}\left[\exists w_{\times} \ s.t.\ d(w_{r})= w_{\times} \right]\\
		&\leq\text{E}\left[\# w_{\times} \ s.t. \ \langle w_{r}, w_{s}\rangle \leq \langle w_{r}, w_{\times}\rangle\right] \\
		&= \sum_{o=0}^{a-1} D_{ w_{s}}\left[ o\right]\text{Pr}\left[d(w_{r})= w_{\times}|o\right]
	\end{aligned}
\end{equation}
where $D_{ w_{s}}$ is the operator that counts the codewords at a distance of $o$ from $w_{s}$. The underlying assumption here is that the overlap between $w_s$ and $w_{\times}$ is uncorrelated among the different $w_{\times}$.

Now all we must do is find $D_{ w_{s}}$. For relatively small values of $a,N$, such codes are known \cite{tableConstWeight} and for larger values we can use the Johnson bounds which apply to $A(N,2u,a)$, the largest size of a constant weight code with parameters $a,N$ and a minimum hamming distance of $2u$  \cite{CodThConstWeight}. Those codes have been studied elsewhere and we will not present them here. Suffice to say for now that, given a code with known upper bounds on $D_{ w_{s}}$ for all overlap values $o$, we can obtain a bound on $p_{\times}$.

\section{Error-free transmission with random constant weight codes}

While designed constant weight codes with a minimum hamming distance have been studied, to the best of our knowledge their random counterpart has not. The use of random codes is justified when the values of $a,N, u$ are too large to make deterministic codes viable, or in the case of neuromorphic sensors where the events that will be observed cannot be predicted a priory and random codes must be considered. In this section we provide the asymptotic results for the operator $D_{ w_{s}}\left[ o\right]$ when the codes are random, again following the logic in \cite{PhyInfComp} for proving the Channel Coding Theorem.

To make this computation, we calculate the probability that a random codeword has an overlap $o$ with the original $w_{s}$ and then the probability that for an overlap of $o$ the errors induce a decoding mistake. Given that the active units are selected with the same probability, we can use combinatorial tools to make such computations.

First, the probability that two random codewords $w_1,\ w_2$ overlap in $o$ active bits is given by
\begin{equation}\label{eq:OverlapProb}
	\begin{aligned}
		\text{Pr}&\left[\langle w_1,w_2\rangle  = o\right]
		= \dfrac{{a \choose o}{N-a \choose a-o}}{{N \choose a}}\\ 
		&\sim \exp\left\{aH\left(\frac{o}{a}\right) + (N-a)H\left(\frac{a-o}{N-a}\right) - NH\left(\frac{a}{N}\right)\right\}\\
		&= \exp\left\{N\left[rH\left(z\right) + (1-r)H\left(\frac{1-z}{1-r}r\right) - H\left(r\right) \right]\right\}
	\end{aligned}
\end{equation}
 and $z=\frac{o}{a}$. Thus the expected number of codewords that overlap in $o$ active units with a random codeword is
 \begin{equation}
 	\begin{aligned}
 		\bar{D}\left[o\right] &= \text{Pr}\left[\langle w_1,w_2\rangle  = o\right] \mathcal{W}\\
 		&\sim \exp\left\{-N \left[
 		\begin{array}{l}
 			rH\left(z\right) + (1-r)H\left(\frac{1-z}{1-r}r\right) \\
 			- H\left(r\right) +\rho
 		\end{array}
 		\right]\right\}
 	\end{aligned}
 \end{equation}
where we define $\rho$ by $|\mathcal{W}| = \exp\left\{\rho N\right\}$. We can now estimate the probability of a wrong decoding
\begin{equation}
\begin{aligned}
	p_{\times} &\leq
	 \sum_{o=0}^{a-1} D_{ w_{s}}\left[ o\right]\text{Pr}\left[d(w_{r})= w_{\times}|o\right]\\
	 &\sim \int_{0}^{1}
	 \exp\left\{-N\left[
	 \begin{array}{l}
	 	\rho +H\left(r\right) 	+ r(1-z)I(p_{10}, p_{01})\\
	 	- rH\left(z\right) - (1-r)H\left(\frac{1-z}{1-r}r\right) 
	 \end{array}\right]\right\}dz,
\end{aligned}
\end{equation}
where $z$ is the variable that integrates over the fraction of overlap. This integral can be approximated by using App.~\ref{App:LaplaceExpand}, thus it becomes
\begin{equation}
	p_{\times} = \text{Pr}\left[d(w_{r})= w_{\times}\right] \sim \exp\left\{N\left[\rho - G(r)\right]\right\}
\end{equation}
where $G(r)$ is defined as 
\begin{equation}
	G(r) = \min_{z\in[0,1]}\left[
	\begin{array}{l}
		H\left(r\right) + r(1-z)I(p_{10}, p_{01}) \\- rH\left(z\right) - (1-r)H\left(\frac{1-z}{1-r}r\right)
	\end{array}
	\right].
\end{equation}

Thus in the limit of large $N$, $|\mathcal{W}| = e^{\rho N}$, the the probability of a wrong decoding is
\begin{equation}
	p_{\times} \sim \exp\left\{N\right[\rho - G(r) \left]\right\}\sim 
	\begin{cases}
		\geq 1\quad \Leftarrow \rho \geq G(r)\\
		0\ \ \ \quad \Leftarrow \rho < G(r)
	\end{cases}
\end{equation}
Hence the transition between almost no decoding errors or almost certainly decoding errors is given by the relative magnitudes of $G(r)$ and $\rho$. We can substitute this new constraint in our cost minimization problem from Eq.~\ref{eq:costMinNoiseless}, which becomes
\begin{equation}
	\argmin_{r,N} C(rN,N) \quad s.t.\ \rho < G\left(r\right),
\end{equation}
which we can solve by setting $N=\frac{\ln |\mathcal{W}|}{G(r)}$ and solving
\begin{equation}
	r^*=\argmin_{r} C\left(r\frac{\ln |\mathcal{W}|}{G(r)}, \frac{\ln |\mathcal{W}|}{G(r)}\right), \quad N^* = \frac{\ln |\mathcal{W}|}{G(r^*)}
\end{equation}

\section{Future work}

This work is still at an early stage and there are a few extensions that will be presented in the future:

\subsection{Precise bounds}

The derivations that we presented give estimates in terms of orders of magnitude, rather than concrete bounds. While the current presentation is easy to understand, more concrete bounds would be necessary. All our derivations are based on Stirling's approximation and Laplace's method for which bounds are well known, therefore this should be a simple computation.

\subsection{Multiple symbols per unit}
In some systems such as neuromorphic cameras there are positive and negative activations \cite{DAVISCamera} and thus it seems natural to expand our results in that direction. The computations are not complicated, as the approximation of the binomial coefficient can easily be extended to the multinomial case (see App.~\ref{App:CombToExp}).

\subsection{Varied codeword probabilities}
In practical cases we are likely to encounter codewords with different probabilities and encoders that cannot accumulate symbols for long enough to let the asymptotic equipartition property alleviate that. Thus, we should consider accounting for probabilities of different symbols and we can add this to our computation by considering decoding as a Bayesian inference problem \cite{InfML} and using the different codeword probabilities as priors.

\subsection{Allowing multiple weights}
If each activation costs energy and some codewords are more frequent than others, it makes sense to allow frequent codewords to have lower weights.

\section{Conclusion}

The derivations presented here give a simple framework to design appropriate fixed weight codes. We believe that our work can be useful in the design of systems for Edge AI, specifically on neuromorphic chips, and which to this day lack theoretical foundations \cite{theoryNeuromorph} or other applications where constant weight codes apply.

\section*{Acknowledgments}

I would like to thank Benjamin Grewe, Stephan Moser, Hui-An Shen, and Jean-Pascal Pfister for valuable informative discussions. P.V.A. was supported by an ETH Postdoctoral Fellowship. 

\appendix

\section*{Approximations for combinatorial expressions}\label{App:CombToExp}
We show here how to approximate some simple terms that appear recurrently in our derivations
\subsection{Binomial Coefficients}
Our main objective here is to show that
\begin{equation}
	{n \choose k} \sim \exp\left[n H\left(\dfrac{k}{n}\right) + O\left(\dfrac{\ln n}{n}\right)\right]
\end{equation}
where $H$ is the entropy function.
This is proven by using Stirling's approximation
\begin{equation}
	\ln n! = n\ln n - n + O(\ln n)
\end{equation}
which can be applied to $n\choose k$,
\begin{equation}
	\begin{aligned}
		\ln {n \choose k} &= \ln n! - \ln k! - \ln (n-k)! \\
		&=  n\ln n - k\ln k - (n-k)\ln (n-k) \\ &+n- k - (n-k)+ O(\ln n) \\
		&= -k\ln\frac{k}{n}-(n-k)\ln\frac{n-k}{n} + O(\ln n)\\
		&=n\left[H\left(\frac{k}{n}\right) + O\left(\frac{\ln n}{n}\right)\right]	
	\end{aligned}
\end{equation}

Finally, note that this can easily be extended to the multinomial case. If we take $k_1,k_2,... k_m$ as the number of symbols of type $1,2,.. m$ subject to $\sum_{i=1}^m k_m = N$,
\begin{equation}
\begin{aligned}
	&{n \choose k_1, k_2, ..., k_m} = \dfrac{n!}{k_1!k_2!...k_m!}\\
	&\sim \exp\left[n H\left(\dfrac{k_1}{n},\dfrac{k_2}{n},...\dfrac{k_m}{n}\right) + O\left(\dfrac{\ln n}{n}\right)\right].
\end{aligned}
\end{equation}

\subsection{Probability of getting k elements out of $n$ if each of the $n$ elements is selected with probability $p$}
\begin{equation}
	{n \choose k} p^k (1-p)^{n-k} = \exp\left[-n \KL\left(\frac{k}{n},p\right) + O\left(\frac{\ln n}{n}\right)\right]
\end{equation}
where $\KL(x,y)$ is the Kullback-Leibler Divergence between Bernoulli distributions with parameters $x$ and $y$. 

From the use of Stirling's approximation on the binomial coefficient we get
\begin{equation}
	\begin{aligned}
		\ln &\left[{n \choose k} p^k (1-p)^{n-k}\right] = -k\ln\frac{k}{n}-(n-k)\ln\frac{n-k}{n} \\
		&+ k\ln p + (n-k)\ln (1-p)	 + O(\ln n) \\
		&=k\ln\left(\dfrac{p}{\frac{k}{n}}\right)+(n-k)\ln\left(\dfrac{1-p}{\frac{n-k}{n}}\right) + O(\ln n) \\
		& = -n \left[\KL\left(\frac{k}{n},p\right) + O\left(\frac{\ln n}{n}\right)\right]
	\end{aligned}
\end{equation}

\section*{A note on Laplace's Method}\label{App:LaplaceExpand}
In our calculations we will use the following approximation
\begin{equation}
	\begin{aligned}
		&\int_{\text{Dom}\left[x_1,x_2,...x_k\right]} \exp\left\{nf\left(x_1,x_2,...x_k\right)\right\}dx_1dx_2...dx_k\\
		&\sim \exp\left\{n\left[f^*+ O\left(\frac{\ln n}{n}\right)\right]\right\}
	\end{aligned}	
\end{equation}
where $n$ is a very large number and $\text{Dom}\left[x_1,x_2,...x_k\right]$ is the domain of integration which must be convex, and contain a single global maximum of the continuous and differentiable function $f(x_1,x_2,...x_k)$ in the integration domain, denoted $f^*$. 

In the case where the maximum of $f$ in the domain of integration is a fixed point the result is a simplification of Laplace's method of integration. We will just add a note for the case where it does not. We will only cover the one dimensional case, noting that the multivariate case requires only a simple but cumbersome modification.

We want to show that if we have a function $f(x)$ on the domain $\left[a,b\right]$ with a single maximum at $f(a)$ that is not a critical point then when $n\rightarrow\infty$,
\begin{equation}
	\int_{a}^{b} \exp\left\{nf(x)\right\}dx \approx  \exp\left\{n\left[f(a) - \frac{\ln(n) + \ln(f' (a))}{n}\right]\right\}.
\end{equation}
The derivation is akin to Laplace's, but on the first order derivative rather than the second order.

In the neighborhood of $a$ we can use the Taylor expansion $f(x) \approx f(a) + f'(a)(x-a)$. Since all points far from $a$ are have $f(x)-f(a)<0$ and this negative value will be scaled by $n$,
\begin{equation}
	\int_{a}^{b} \exp\left\{nf(x)\right\}dx \approx  \int_{a}^{b} \exp\left\{n\left[f(a) + f'(a)(x-a)\right]\right\}dx.
\end{equation}
Finally, since the exponential decay is also scaled by $n$, all values of $b$ away from $a$ will give similar results
\begin{equation}
\begin{aligned}
	\int_{a}^{b} &\exp\left\{n\left[f(a) + f'(a)(x-a)\right]\right\}dx\\
	&\approx \exp\left\{nf(a)\right\}\int_{a}^{\infty} \exp\left\{nf'(a)(x-a)\right\}dx
\end{aligned}
\end{equation}
and by evaluating the last integral we get 
\begin{equation}
	\int_{a}^{b} \exp\left\{n\left[f(a) + f'(a)(x-a)\right]\right\}dx
	\approx \dfrac{1}{nf'(a)}\exp\left\{nf(a)\right\}
\end{equation}




\end{document}